\documentclass[superscriptaddress, aps, prl, reprint, twocolumn, amsmath, amssymb, showpacs]{revtex4-1}
\usepackage{graphicx}  % needed for figures
\usepackage{mathptmx}
\usepackage{epstopdf}
\usepackage{dcolumn}   % needed for some tables
\usepackage{bm}        % for math
\usepackage{amssymb}   % for math
\usepackage{amsmath}   % for matrix
\usepackage{amsthm}
\usepackage{chemarrow}
\usepackage{color}
\usepackage{xcolor} 
\usepackage{mathrsfs}
\usepackage{float}
\usepackage{physics}
\usepackage{bbold}
\usepackage{bbm}
\usepackage[toc,page]{appendix}
\usepackage{longtable}
\usepackage{multirow}
\usepackage{upgreek}
\usepackage{epstopdf}
\usepackage{ulem}
\usepackage{graphicx}
\usepackage{lipsum}

\usepackage[a4paper, colorlinks=true, linkcolor=blue, citecolor=blue,
pdfauthor={ },
pdftitle={ },
pdfsubject={ },
pdfkeywords={ }]{hyperref}
\usepackage[version=4]{mhchem}
\usepackage{ulem,fancyvrb}

\theoremstyle{plain}
\newtheorem*{theorem*}{Theorem}

\begin{document}
	
	%\preprint{APS/123-QED}

\title{Levitated Sensor for Magnetometry in Ambient Environment}

%\footnotetext[1]{These authors contributed equally.}
 % You can write out first names or use initials - either way is acceptable, but be consistent
 \author{
    Wei Ji$^{1,2,3\ast\dagger}$,
    Changhao Xu$^{2,3\dagger}$,
    Guofeng Qu$^{4}$, Dmitry Budker$^{2,3,5,6}$\\
    % Additional lines of authors should be inserted using the \and command (not \\)
    % Institution list, in a slightly smaller font
    \small$^{1}$School of Physics and State Key Laboratory of Nuclear Physics and Technology,  Peking University, Beijing, 100871, China. \\
    \small$^{2}$Johannes Gutenberg-Universit{\"a}t Mainz, 55128 Mainz, Germany.\\
    \small$^{3}$Helmholtz Institute Mainz, 55099 Mainz, Germany.\\
    \small$^{4}$Key Laboratory of Radiation Physics and Technology of the Ministry of Education, \small Institute of Nuclear Science and Technology, Sichuan University, 610064 Chengdu, China.\\
    \small$^{5}$GSI Helmholtzzentrum für Schwerionenforschung GmbH, 64291 Darmstadt, Germany.\\
    \small$^{6}$Department of Physics, University of California, Berkeley, CA 94720-7300, USA.\\
    % Identify at least one corresponding author, with contact email address
    \small$^\ast$Corresponding author. Email: wei.ji@pku.edu.cn
    % Joint contributions can be indicated like this
    \small$^\dagger$These authors contributed equally to this work.
    }
%\date{\today}

\begin{abstract} 
Levitated particle systems have gained significant attention as a rapidly advancing platform for precision sensing, offering low-loss, highly isolated environments by eliminating mechanical contact and associated noise. Current room-temperature levitation techniques are primarily sensitive to acceleration, with magnetic sensing often relying on the Meissner effect, which is impractical under ambient conditions. Here, we demonstrate a diamagnetically stabilized magnetically levitated magnet magnetometer (LeMaMa), where the motion of the magnet is detected optically. Leveraging strong spin-lattice coupling in the ferromagnet to suppress spin-projection noise and minimizing dissipation through levitation, we achieve a sensitivity of 32\,fT/$\rm\sqrt{Hz}$. This sensitivity is adequate for a wide range of applications in biology, chemistry, and fundamental physics, matching the performance of leading technologies like SQUIDs and atomic magnetometers, while offering the distinct advantage of operating at room temperature and under Earth’s magnetic field. 
\end{abstract}
\maketitle

\noindent Levitating nano- and micro-objects in vacuum is a rapidly developing field with tremendous potential\,\cite{gonzalez2021levitodynamics}. The absence of physical clamping minimizes interaction with the environment and enables tunable performance through precise control of the trapping potential\,\cite{tebbenjohanns2021quantum,rieser2022tunable,magrini2021real}. This results in a system with low loss and effective isolation from environmental noise, making it ideal for high-precision sensing applications. For example, such systems can achieve unprecedented sensitivity in acceleration, force and rotation detection\,\cite{ranjit2016zeptonewton,hempston2017force,moore2014search,ricci2019accurate,leng2024measurement,ahn2020ultrasensitive,wang2020diamagnetic}. Moreover, if a levitated system can be made sensitive to electromagnetic signals, its range of applications would expand significantly. Sensitive magnetometers are essential for detecting faint magnetic signals from human brain and heart \cite{boto2018moving, cohen1972magnetoencephalography, aslam2023quantum} and for probing dark matter and physics beyond the Standard Model \cite{lee2023laboratory, safronova2018search, wei2023ultrasensitive}. To achieve magnetic field sensitivity, the levitated object must exhibit magnetic properties, such as magnetic moments or induced currents, enabling it to interact with external magnetic fields effectively\,\cite{delord2020spin,hofer2023high,ahrens2024levitated}.

Electron-spin-based magnetometers, including spin exchange relaxation-free (SERF) atomic magnetometers, are extensively utilized for precision magnetic field measurements\,\cite{kominis2003subfemtotesla,savukov2005tunable,wei2023ultrasensitive}. Recent advancements, such as the integration of quantum entanglement, are expected to significantly enhance the magnetic sensing capabilities \cite{block2024scalable}. It was also suggested that even without entanglement, ferromagnetic materials can achieve a magnetic sensitivity that far exceeds the standard quantum limit for independent spins \cite{kimball2016precessing,vinante2021surpassing}. The core advantage of this approach lies in the correlations of spins within the lattice structure of the magnet, which reduces spin projection noise, a key limitation in spin-based sensors like atomic magnetometers\,\cite{budker2007optical}. The main challenge lies in achieving stable and convenient levitation in a practical and efficient manner without introducing noise from the levitation mechanism itself. 

One approach to enhancing sensitivity is operating in a cryogenic environment, as lower temperatures generally reduce thermal noise. Building on this approach, a sensitivity of $\text{370\,pT}/\sqrt{\text{Hz}}$ was achieved using a levitated ferromagnetic object with an attached mirror for optical detection\cite{jiang2020superconducting}. In a related study, a superconducting levitated magnet sensor achieved a sensitivity of $\text{20\,fT}/\sqrt{\text{Hz}}$, utilizing a high-sensitivity superconducting quantum interference device (SQUID) magnetometer as the probe\,\cite{ahrens2024levitated}.
However, these techniques require a cryogenic system, which adds significant operational complexity and practical limitations. Various important applications, particularly the detection of biomagnetic signals require ambient conditions\,\cite{shi2015single,degen2009nanoscale,cohen1972magnetoencephalography,boto2018moving,aslam2023quantum}. Operating at room temperature offers a more feasible and accessible solution but shifts the primary challenge to achieving stable levitation and suppressing dominant noise through non-thermal methods. 

To address this, we present a novel room-temperature diamagnetically stabilized levitated magnet magnetometer (LeMaMa). We levitate the magnet via the magnetic gradient field of a lifting magnet and stabilize the levitation with a single layer of diamagnetic material \cite{simon2001diamagnetically}. We use optical reflection to monitor the dynamics of the magnet. To suppress diverse noise sources, we implement advanced methodologies that integrate innovative structural design and material engineering, including a precisely engineered diamagnetic layer to mitigate dominant thermal noise from conducting materials. Operating at an optimal frequency of approximately 305\,Hz, our setup achieves a sensitivity of 32\,fT/$\sqrt{\text{Hz}}$ with a 0.2\,mm$^3$ magnet, representing a several-orders-of-magnitude improvement over the previously demonstrated reflection magnetometer based on superconducting levitation \cite{jiang2020superconducting}. Furthermore, this sensor operates in an ambient environment at room temperature, tolerates residual magnetic fields greater than Earth’s field, and outperforms comparably sized leading SERF and SQUID magnetometers in terms of environmental adaptability. These features make it suitable for a wide range of applications, including biological sensing, chemical sensing, and fundamental physics research.

\begin{figure}
    \centering
\includegraphics[width=0.9\linewidth]{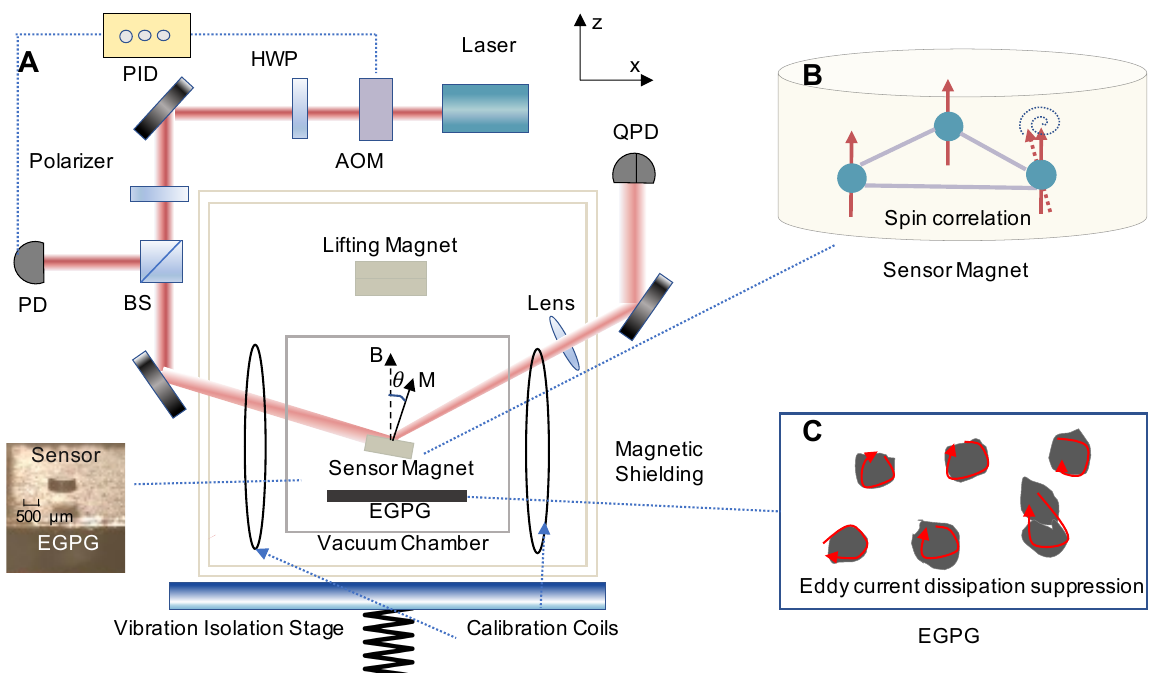}   \caption{\textbf{Experimental Setup.} (A) The sensor magnet\,(SM) is levitated with a lifting magnet.  A layer of epoxy glued pyrolytic graphite (EGPG) is placed below the SM to stabilize the levitation. A photo of the SM and the PG is shown in the inset. A position sensitive quadrant photodiode detector (QPD) is used to observe the reflected light beam from the sensor. The laser power is stabilized with a proportional-integral-derivative (PID) system with the input of a reference probe channel and a control unit composed of an acousto-optic modulator (AOM). HWP: half wave plate, BS: beam splitter.(B) A diagram illustrating spin correlation and the response to random spin noise. The spins are strongly correlated along the z-axis, and any disturbance on individual spin is rapidly relaxed due to the Gilbert relaxation. (C) A diagram of the EGPG plate, illustrating the distribution of pyrolytic graphite particulates. The eddy currents (red curves) are confined to the regions within the particulates, significantly reducing the dissipation rate compared to that of a large eddy current loop in normal pyrolytic graphite.}
    \label{fig1:structure}
\end{figure}

\subsection*{Experimental Setup}

The LeMaMa setup is illustrated in Fig.\,\ref{fig1:structure}. The sensor magnet (SM) has the shape of a disk with a radius of $410\pm2\,\mu$m and a thickness of $380\pm2\,\mu$m, with polished upper and lower surfaces for improved reflectivity. The lifting magnet (LM), responsible for generating the magnetic force that counteracts the gravitational pull on the SM, consists of a stack of 80 disk magnets, each 10\,mm in diameter and 0.5\,mm in thickness. A 1-mm-thick layer of diamagnetic material, epoxy-glued pyrolytic graphite (EGPG), is placed beneath the SM to provide magnetic stabilization via its induced magnetic field \cite{simon2001diamagnetically}. This EGPG primarily consists of pyrolytic graphite, which is ground into powder and bonded with epoxy to minimize eddy current dissipation \cite{chen2022diamagnetic} and the Johnson magnetic noise it produces.
%For clarity, we denote this material as PG in the figure.
The interplay between gravitational force, the magnetic force from the LM, and the repulsive force from the EGPG creates a local potential minimum, allowing the sensor magnet to be stably levitated\,\cite{simon2001diamagnetically}.

The dynamics of the sensor magnet are detected using an optical setup. A laser beam with a wavelength of 852\,nm is incident onto the SM at a grazing angle in the \textit{x-z} plane. 
 To maximize reflectivity and minimize absorption, we chose the light to be s-polarized. 
Laser light reflected from the surface of the SM is collected with a position-sensitive quadrant photodiode (QPD) detector. The rotational motion of the magnet in the incidence plane is magnified on the detector due to a 3.4\,m long optical lever and amplifying the sensitivity of the sensor to a magnetic field along the \textit{x}-axis. In contrast,  translational modes are not amplified, effectively suppressing crosstalk induced by vibrational noise.

The SM and EGPG stabilizer are housed within a glass vacuum chamber that is evacuated and sealed with a vacuum valve, then disconnected from the vacuum system to minimize vibrational noise. In a typical measurement condition, the residual pressure is 0.025\,mbar. The chamber is subsequently relocated to the detection area, where it is shielded within a four-layer cylindrical magnetic enclosure to ensure environmental isolation. To further suppress vibrational interference, the entire setup, except for the QPD and laser source, is mounted on a passive vibration isolation stage, allowing for an extended optical lever arm and minimizing disturbances from the environment through the connection cables.

\subsection*{Dynamics of spin and mechanical system}
The correlation of spins in a ferromagnet arises from their exchange interaction in the lattice field. Unlike atomic magnetometers, which require polarizing electron spins optically, spins in magnetized ferromagnetic systems are already highly polarized at room temperature. This intrinsic polarization offers significant advantages: the spin density (on the order of $ 10^{22} \hbar/\rm cm^3$) is approximately seven orders of magnitude  higher than in atomic vapor cells (on the order of $ 10^{15} \hbar/\rm cm^3$ for Rb when heated to 200$^\circ$C), reducing spin-projection noise and improving the signal-to-noise ratio\,\cite{kimball2016precessing}. Furthermore, the strong spin-lattice coupling facilitates rapid averaging of spin projection noise. As illustrated in Fig.\ref{fig1:structure}(b), the disturbance to specific spin relaxes within nanoseconds because of the Gilbert damping. Since detection is at much lower frequencies, this leads to enhanced sensitivity for low-frequency magnetic field measurements \cite{kimball2016precessing}. These combined benefits—simplified polarization, higher spin count, and intrinsic noise suppression—make ferromagnetic spin systems highly effective magnetic field sensors.

The principle of suppressing eddy current dissipation is illustrated in Fig.\,\ref{fig1:structure}(c). When normal pyrolytic graphite is used, eddy currents flow through the entire region, forming large current loops. In contrast, with isolated particles, the eddy currents are confined to small loops, as the eddy current dissipation per unit volume scales quadratically with particle size, smaller area significantly reducing dissipation\,\cite{jiang2020superconducting}, and enabling higher quality factors.
%\subsection{Sensitivity}

 The resonance frequency of the sensor torsional mode is $f_0=\sqrt{MB/I}/2\pi$, where  $B$  is the average magnetic field applied on the sensor magnet, $M$ is the magnetic moment of the sensor and $I$ is its appropriate moment of inertia. In a typical case, the magnetic field  $B_{0}= 1.92(3)\,\rm mT$  is primarily along \textit{z}-axis which is produced by the lifting magnet, resulting in a resonance frequency of  $f_0 = 304.84\,\rm Hz$. We focus on the torsional mode in the \textit{x-z} plane where the sensor is most sensitive to magnetic field along the \textit{x}-axis. 

\begin{figure}
    \centering \includegraphics[width=0.5\columnwidth]{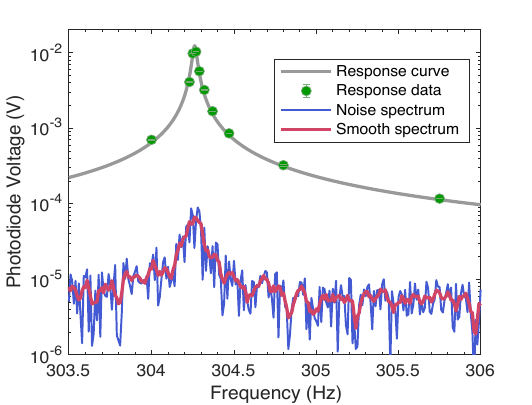}
    \caption{ \textbf{The response and noise root-mean-square spectra around the response frequency.} The green dots represent the response of the system to a 10\,pT amplitude magnetic field under different driving frequencies and the gray line is the corresponding fit. The voltage noise spectrum of the sensor is shown in blue line, with the red curve being a smoothed version of it.}
    \label{fig2: response-and-sensitivity}
\end{figure}

To characterize the response of the magnetometer, we apply an alternating magnetic field along the \textit{x}-axis with an amplitude of  $10 \,\mathrm{pT}$ using a set of field coils placed inside the magnetic shielding. By varying the frequency of the applied magnetic field, we record the corresponding root-mean-square (RMS) of the output voltage to evaluate the magnetometer performance. The response is fitted using the function  $A/\sqrt{(f^2 - f_0^2)^2 + \gamma^2 f^2}$ to obtain the response curve. The fitted response within the near-resonance region is shown in Fig.\,\ref{fig2: response-and-sensitivity}(A). The full-width-half-maximum of the response curve is approximately 0.025\,Hz, corresponding to a quality factor of about 12000. 
%and a dissipation rate of $\gamma \approx 0.025\,\mathrm{Hz}$.
This indicates that the system operates in a low-dissipation mode, making it ideal for high-sensitivity measurements. For reference, we also present the root-mean-square voltage spectrum obtained by a Fourier transform of the time-dependent signal with no magnetic field applied (blue and red curves). The response curve and noise spectrum over the full frequency range is depicted in Supplementary Text.

\begin{figure*}
    \centering \includegraphics[width=1.05\linewidth]{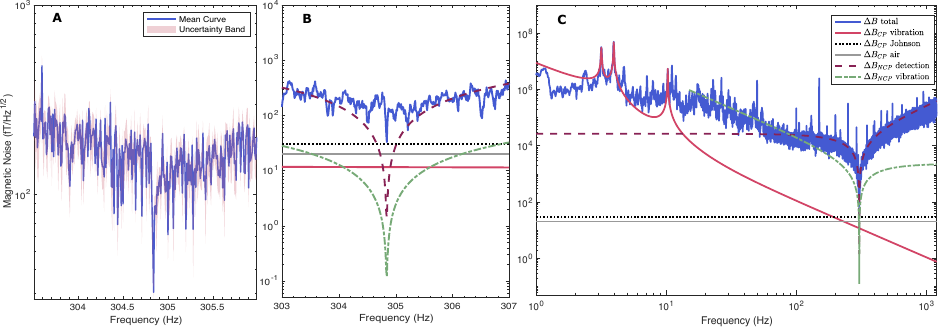}
    \caption{ \textbf{Magnetic sensitivity across different frequency ranges.} (A) The blue line represents the average value across five datasets, while the pink band indicates the corresponding uncertainty for each frequency bin. (B) and (C) show the sensor’s noise spectrum, highlighting contributions from different noise sources. Panel (B) focuses on the near-resonance region, while panel (C) covers a broader frequency range. In both plots, the blue line represents the total noise, corresponding to the sensor’s sensitivity. Distinct peaks are observed around 3\,Hz, 4\,Hz, and 10\,Hz, corresponding to the translational modes along the $x, y$ and $z$ axes, respectively. The $\rm N_{CP}$ noise includes magnetic noise from the conducting material, shown by the black-dashed line, as well as vibrational noise represented by the red line (rescaled from the actual noise background) and the yellow line (determined by fitting the vibration noise curve). The solid gray line indicates the air collision noise. The $\rm N_{NCP}$ noises are shown with the purple-dashed curve the vibrational noise from the optics not on the vibration stage and green-dashed line detection noise from the data acquisition system.}
    \label{fig3:sensitivity-noise-analysis}
\end{figure*}

The voltage noise spectrum of the sensor can be converted into a magnetic field noise spectrum by dividing it by the sensor response factor curve, which relates the voltage output to the magnetic field input. The resulting magnetic noise spectral density for near resonance range is presented in Fig.\,\ref{fig3:sensitivity-noise-analysis}(A). We collect five datasets, each 100 seconds long, perform a Fourier transform on each, and then calculate the average and the uncertainty for each frequency bin. The sensitivity at the on-resonance frequency is $32 \pm 3 \, \mathrm{fT/\sqrt{Hz}}$.

\subsection*{System sensitivity studies}

The noise spectrum, along with an assessment of various noise sources, is shown in Fig.\,\ref{fig3:sensitivity-noise-analysis} (B) and (C). The blue line represents the total noise spectrum of the sensor. Noise can broadly be classified into two categories: coupled noise ($\mathrm{N_{CP}}$), which directly influences the sensor dynamics, and non-coupled noise ($\mathrm{N_{NCP}}$), which does not. For the near-resonance range, the non-coupled noise $\rm \Delta B_{NCP}$ typically exhibits a dip in the curve. This occurs because the original noise, which is generally flat over this range, is divided by the response curve that has a pronounced peak. In contrast, the converted coupled noise $\rm \Delta B_{CP}$ remains relatively flat since its response in the voltage spectrum is already largely shaped by the response curve (see the peak in the voltage noise spectrum in Fig.\,\ref{fig2: response-and-sensitivity}), consistent with the fluctuation-dissipation theorem. A detailed analysis of these noises are explained in the Supplementary Text.

The dominant noise near resonance is magnetic noise originating from Johnson noise induced by thermal currents in surrounding conducting materials\,\cite{lee2008calculation}, particularly from the diamagnetic plate and the metal vacuum chamber. To diminish this noise, we replace the aluminum vacuum chamber with glass one and mesh normal pyrolytic graphite into powder and use epoxy to isolate the particulates and form a new plate. This effectively reduces the thermal current magnetic noise from  $110\,\rm{fT/\sqrt{Hz}}$ to $30\,\rm{fT/\sqrt{Hz}}$ (black dotted line). This noise is validated by measurements using a commercial atomic magnetometer. A secondary noise source is air collisional noise from gas atoms. Using the Monte Carlo method, we simulate this noise at the working condition with a pressure of 0.025\,mbar and a room temperature of 293\,K, estimating it to be below $20\,\rm{fT/\sqrt{Hz}}$ (solid black line). 

The system also experiences vibrational noise, primarily from seismic noise, along with air convection and acoustic noise within the laboratory. The coupled type vibrational noise drives the support system of the sensor and subsequently affect the dynamics of the sensor. The 3\,Hz, 5\,Hz and 15\,Hz peaks in the blue line are the noise-driven translational mode along $x,y,z$-axis respectively. We use the response curves of the vibrational mode to estimate the noises, depicted as the solid purple line. The noise level is consistent with the assessment obtained through an alternative approach, utilizing vibrational data and the transmission factor of the vibration isolation stage. These estimations give us $10\,\rm{fT/\sqrt{Hz}}$ at the resonance frequency. The vibrational noise from the photodetector system, which is mechanically isolated from the sensor support structure, represents a non-coupled noise source (depicted by the green dashed line). This noise dominates in the frequency range from 15 to 80 Hz and is estimated to be approximately $0.1 \,\mathrm{fT/\sqrt{Hz}}$ at the on-resonance frequency. Other main contributors to non-coupled noise include the photon detection noise, which consists of photonic noise from the photodetector and electronic noise from the entire electronic system (purple-dashed line).  It is estimated to be around $2 \,\rm{fT/\sqrt{Hz}}$.

As shown in Fig.\,\ref{fig3:sensitivity-noise-analysis}(C), the sensor sensitivity ranges from approximately 1\,$\rm nT/\sqrt{Hz}$ at low frequencies below 15\,Hz to approximately $\rm 200 \,pT/\sqrt{Hz}$ in the frequency range between 80\,Hz and 200\,Hz.
 Noise from the power supply at 50 Hz and its harmonics appears as distinct spikes in the spectrum. 
 
\begin{figure}
    \centering \includegraphics[width=0.5\linewidth]{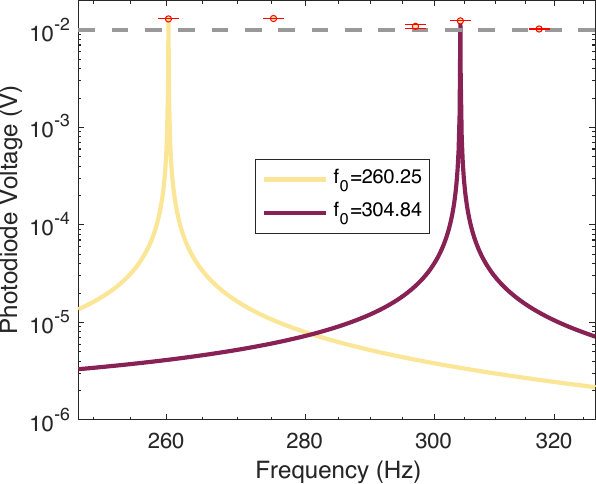}
    \caption{\textbf{On-resonance response for several values of resonance frequency.} We tune the resonance frequency from 260\,Hz to 318\,Hz by changing the leading field $B_z$. A 10\,pT amplitude magnetic field is applied to drive the magnet oscillation on resonance. The red dots represent the amplitude of the response and the yellow and purple lines represent the frequency response curve of the corresponding resonance conditions. The dashed line at $10^{-2}$\,V is shown to guide the eye.}
    \label{fig4:tune}
\end{figure}

The resonant sensor with ultrahigh sensitivity at resonance frequencies can be highly useful for applications with signals modulated to be on resonance with the sensors, for instance the detection of the exotic interactions beyond the standard model of particle physics\,\cite{wei2022constraints}. If the resonance frequency can be dynamically tuned\,\cite{savukov2005tunable}, the system can operate as a closed-loop magnetometer \cite{etde_21420025,yan2022three}, significantly extending both its frequency response range and dynamic range. As shown in Fig.\,\ref{fig4:tune}, we demonstrate the ability to adjust the resonance frequency by varying the leading magnetic field using a set of coils generating a uniform field along the \textit{z}-axis. By tuning the magnetic field  $B_z$  from 1.42\,mT to 2.12\,mT, we successfully shift the resonance frequency from 260\,Hz to 318\,Hz. The corresponding on-resonance responses are represented by the red points. Additionally, two response curves for near resonance modes, with resonance frequencies of 260.25\,Hz and 304.84\,Hz are also depicted.  %These results show the capability to operate the sensor as a tunable resonant magnetometer over a broad frequency range.

\subsection*{Discussion}

Our LeMaMa magnetometer demonstrates a sensitivity comparable to the leading magnetometry technologies like SERF\cite{kitching2018chip} and SQUID\,\cite{fong2005high} of similar dimensions. However, SERF magnetometers require elevated temperatures and can only operate in sub-microtesla fields (much smaller than the Earth field of 20-50\,$\rm \mu$T), while SQUIDs need cryogenic conditions and are not trivial to operate in the Earth field. In practice, these sensors are typically used in shielded environments, adding complexity, challenging biological applications, and increasing the distance between the sensor and test sample. This distance limits spatial resolution and sensitivity to signals of interest. Note that the magnetic shielding in our setup is used to suppress alternating magnetic noise in the environment, such as 50\,Hz power line noise. The magnetometer is capable of operating directly in ambient conditions and can function in higher magnetic fields, up to the millitesla range, as demonstrated here. Our setup achieves a sensitivity several orders of magnitude higher than that of a superconducting levitated mirror \cite{jiang2020superconducting}, primarily due to improved control over magnetic and vibrational noise. 
%The ease of room-temperature operation facilitates optimization. 
While superconducting levitation offers lower thermal noise, it is typically not the dominant noise source, especially for sensors larger than 100 micrometers \cite{vinante2021surpassing}. 

The LeMaMa sensor can be used for fundamental physics research, including searching for exotic spin-dependent forces and dark matter \cite{ahrens2024levitated,fadeev2021ferromagnetic,vinante2020ultralow}. Unlike atomic magnetometers that have heating components and glass cell window that limits the distance between the exotic force source and the sensor, the source can be placed just beside the LeMaMa sensor magnet and a force range below milimeter and mediator axion mass higher than 10\,$\mu$eV is achievable\cite{cong2024spin}, which is within the well motivated axion window\cite{borsanyi2016calculation,youdin1996limits}. The sensor can also be utilized for biomagnetic signal detection in ambient environments\,\cite{shi2015single,cohen1972magnetoencephalography,boto2018moving,aslam2023quantum}, with the advantage of being positioned closer to the sample, resulting in a stronger signal.

A straightforward approach to improving the sensitivity of the resonance mode is to reduce the dominant magnetic noise from surrounding materials, especially the EGPG. Replacing the EGPG with isolating diamagnetic materials is one solution; however, it is crucial to select materials with the highest possible diamagnetic susceptibility for the benefits of higher levitation height\cite{simon2001diamagnetically}. Air collisional noise can be mitigated by achieving a higher vacuum, but this introduces a trade-off with heating. A more effective solution is to enhance the reflectivity of the magnet using high-reflection coatings and employing higher-sensitivity photodiodes. With these improvements, a sensitivity of 0.1\,$\rm{fT}/\sqrt{Hz}$(vibrational noise limited) is achievable. For the broadband mode, the dominant noise below 100\,Hz originates from vibrational vibrations. Performance can be improved by reducing vibrational noise, such as relocating the sensor to low-noise laboratories\,\cite{ze2001seismic} or doing an experiment in space\,\cite{fadeev2021gravity}.

\noindent

\bibliographystyle{utphys}
%\bibliography{mainrefs}

\section*{Acknowledgments}
The authors thank D.F.J. Kimball, H. Ulbricht and A. Vinante for helpful conversations.
W. J. is funded by the startup grant of Peking University. W. J.,  C. X., and D. B. are funded by the QuantERA project
LEMAQUME (DFG Project No. 500314265) and by the DFG Project ID 390831469: EXC 2118 (PRISMA+ Cluster of Excellence).

%%%%%%%%%%%%%%%% REFERENCES %%%%%%%%%%%%%%%

%\clearpage 
%\bibliography{LeMaMa} 

\bibliographystyle{sciencemag}
\end{document}